# Integrating Heterogeneous Building and Periphery Data Models at the District Level: The NIM Approach

Timo Greifenberg, Markus Look, Bernhard Rumpe
*Software Engineering, RWTH Aachen University, http://www.se-rwth.de/*

Keith A. Ellis
*Intel Labs Europe, Intel Ireland Ltd., http://www.intel.eu/labs*

ABSTRACT: Integrating existing heterogeneous data models for buildings, neighbourhoods and periphery devices into a common data model that can be used by all participants, such as users, services or sensors is a cumbersome task. Usually new extended standards emerge or ontologies are used to define mappings between concrete data models. Within the COOPERaTE project a neighbourhood information model (NIM) has been developed to address interoperability and allow for various kinds of data to be stored and exchanged. The implementation of the NIM follows a meta model based approach, allowing for runtime extension and for easily integrating heterogeneous data models via a mapping DSL and code generation of adaptation components.

## 1 INRODUCTION

The COOPERaTE (COOPERaTE, 2013) project is challenged with developing and demonstrating an open, scalable neighbourhood service and management platform that integrates local monitoring and control functions within the built environment with the power, flexibility and scalability of cloud computing for the delivery of energy and other valued services based on innovative business models.

To realise that challenge the COOPERaTE platform must acquire, aggregate, analyse and act on data from disparate and dispersed sources. The requirement for interoperability between these ICT sources is paramount and depends on high-level data models where standardisation is more easily attainable.

Such data models are needed to enable data exchange and interaction between different ICT tools and are thus needed for value-added services aiming at increasing the energy efficiency of the overall neighbourhood. The COOPERaTE data model - the neighbourhood information model (NIM) - has been developed to address interoperability and allow for various kinds of data to be stored and exchanged:
- time series data such as measured sensor data, which changes in real-time
- data describing the installed systems, which changes infrequently
- constant data, which never changes, such as the geographical position of a building.

Additionally such data might be stored as factual data (historical and real-time measurements) and forecast data, acquired by external forecast services or through simulation and optimization services. Typically each data entry also contains additional information, such as the last change, its time of creation or possible value ranges.

A NIM must be capable of integrating all data already available through either an existing building information model (BIM) or building energy management systems (BEMS) using heterogeneous standards. Even for a single concrete neighbourhood, integration is complex and has to define which information is semantically equal within the different standards. One approach to solve this problem is the use of ontologies and mappings between them. Within the eeSemantics Community (eeSemantics, 2012) there has been a lot of research in defining ontologies for buildings and neighbourhoods at different scales. Such an ontology typically results in a new high level data model which has to be extended manually if new requirements arise or new data models have to be integrated. Integrating the wide range of existing data models within a single ontology is infeasible due to the myriad of existing models.

This paper presents an alternative approach to defining a NIM where data models can be integrated by dynamically generating adapters between data models. The NIM is integrated in a Neighbourhood Energy Management System used within the COOPERaTE project, enabling value-added services. The implementation of the data model follows a meta model based approach, such as in the Internet-of-Things approach, allowing for runtime extension of the overall system and for easily integrating het-erogeneous data models via adaptation by using formal models. As a frontend the SEMANCO data model (Corrado & Ballarini, 2012) is extended serving as a basic ontology for new services. This paper presents an approach to creating simple, reusable models of existing data models already used in a neighbourhood for integrating the data into the overall NIM. As a language for expressing the models to be integrated we use a Domain Specific Language (DSL) defined by a context free grammar. DSLs are widely used within model based software engineering and enable the domain expert to express knowledge in a language specifically designed for the domain. Such a language provides domain specific concepts as first-level language concepts.

By having such a formal description for each used data model the complete neighbourhood can be integrated into the NIM. Thus, value-added services using the overall neighbourhood data can be provided. Based on these models, code generation techniques are used to generate adapters out of the abstract mapping description. By adding a new model to the Neighbourhood Energy Management System, the NIM is extended at runtime and a new adapter is generated which allows access to the data fitting to the new model. Thus a specific NIM for a neighbourhood can be implemented quickly by defining models or even reusing existing models resulting in faster integration of new data sources at runtime. The exchangeability and reuse of mapping models can additionally be increased by composing them into libraries. Apart from the integration of existing data models the implementation of new value-added services within the Neighbourhood Energy Management System follows a similar approach shown in this paper.

The rest of the paper is structured as follows: Section 2 will describe the necessary background, Section 3 will introduce the NIM, Section 4 will provide related work and Section 5 will give a conclusion and an outlook to future work.

## 2 BACKGROUND

In this section we will provide the necessary background for our approach which relies heavily on model based and generative software engineering instruments and techniques.

One of these instruments are DSLs (Fowler, 2010) which can be used to enable domain experts to express their knowledge with first class language concepts. By using a DSL an abstraction from the technological problem space to the domain problem space can be achieved. For defining DSLs we use our framework MontiCore (Krahn, et al., 2010) that uses context free grammars as language definitions and is able to provide a parser, a prettyprinter, editors, context condition support and support for code generation (Völkel, 2011). For implementing code generators we use the template engine freemarker. Some languages provided by MontiCore include the UML/P (Rumpe, 2012),(Rumpe, 2011), (Schindler, 2012), a slightly modified derivate of the UML (OMG, 2011).

A code generator in principal provides a transformation between the model, defined in the DSL, and the specific source code required for a certain technology. Thus the code generator is able to transform the model into source code, which is Java in our case, and to add the required technological specifica.

Even though the effort for designing a DSL and implementing a code generator is typically high, additional value is created by reusing the generator for different scenarios with different models. By following certain guidelines (Karsai, et al., 2009) the effort for creating a DSL and a code generator can be lessened. For our approach we use MontiCore to design a DSL that is able to express a mapping between two data models, to check context conditions and generate specific source code out of the models.

## 3 THE NEIGHBOURHOOD INFORMATION MODEL

As part of the COOPERaTE Project a concrete Data Model for Neighbourhoods, the Neighbourhood Information Model (NIM), was developed. The purpose of this model is to gain a common understanding on the data needed for describing the neighbourhood in a way that all the value-added services planned to be developed within the project can be implemented. To achieve this goal the NIM is based on two different information sources: The identified requirements and use cases of the project (Pesch, et al., 2013) and the already existing data models of the building domain (Corrado & Ballarini, 2012). For the implementation of a prototype we chose a flexible approach which goes beyond simply using the resulting model as domain model. The reason is that we consider a flexible platform where the data model can be extended as described in the introduction as needed.

Within this section we first present the concrete data model that defines available data fields within our system. After that we present the meta model which can be seen as an abstraction of the concrete model which is used as domain model of the prototype. After that we introduce our concept of defining the data model in a DSL and the generation of required services for accessing the data.

### 3.1 *A Concrete Neighbourhood Information Model*

For a concrete data model we reused and extended already existing data models. The SEMANCO data model was extended to incorporate the notion of a neighbourhood by including additional categories. The concrete entries and the details of the data model can be found in Deliverable D1.2 (Look & Greifenberg, 2013) of the COOPERaTE Project. As shown in Figure 1 the neighbourhood element serves as a composite for the other elements. A neighbourhood itself can contain information on traffic, persons, reports, geographical information, energy grid connections and other elements. Shown by the subclass hierarchy elements can be parking spaces or energy elements which can contain information on energy data and are divided into public lighting, building, technical systems and electric vehicles. Apart from the extension by a neighbourhood element we introduced the information on the energy grid connection, which has links to exactly two energy elements. This data field is required to store information about energy connections within the neighbourhood. This is especially needed when talking about other scenarios then electrical since there is not always an underlying grid present.

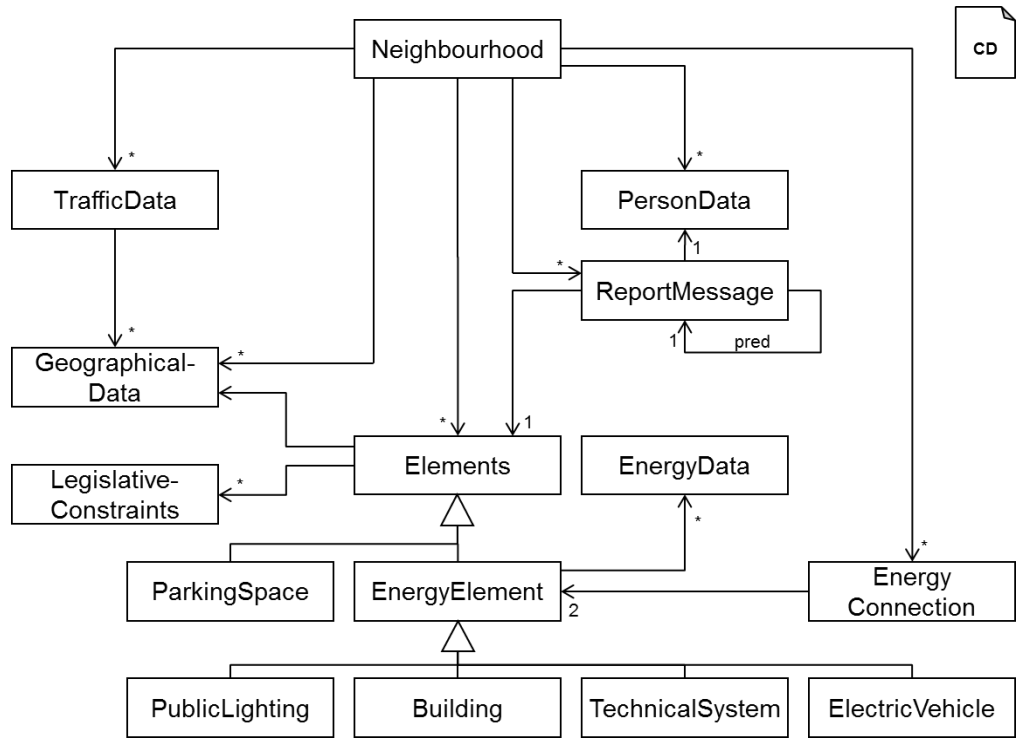

Figure 1. The extended concrete data model used within COOPERaTE. Based on the SEMANCO data model new categories such as a top level neighbourhood, as well as parking spaces and public lighting have been introduced.

### 3.2 The Generic Meta Model

To handle the huge variety of heterogeneous data entries from several existing data models and to enable integration, a generic meta model of the concrete NIM was developed.

As shown in Figure 2 the NIM consists of several NIMComponents. These components follow the composite design pattern (Gamma, et al., 1994) and are either categories or entries where the categories in turn can contain other components. The categories are linked to each other via an association to enable cross referencing between categories.

Moreover the entries contain information about their metadata such as a name, a unit and privacy enabling fields. In the context of each entry, different kinds of value data can be stored. The kinds of values taken into account are: values, value ranges, forecast data, historical data, meta information.

Values represent the measured, calculated or manually inserted information of an entry. Each value has a timestamp whereby the value with the most recent timestamp is considered as the actual value of an entry. In addition to the actual value the older values are of course available and thus historical information for an entry is available. The value itself is stored in the value field within the value element. Forecast data is data which is assumed to be present in the future at a specific point in time. The forecast data is modelled as an explicit element in contrast to the historical data which is model implicitly via the timestamp. This explicit modelling is necessary since several predictions for the same future, the same point in time, should be possible. Thus for each prediction a forecast is created that contains values for a certain period of time. To distinguish between different forecast sources a source identifier is contained within the forecast element. The value range stores the upper and lower bound of valid values.

Additionally, security aspects have been directly woven into the data model to ensure that no data is stored if a user, service or system does not consent with storing. The addition of an expiry date for each value can be used to delete stored values after a certain point in time. Since historical data can also be queried via the data model this is especially helpful, because it enables that values can be padded automatically and that users or systems can specify the lifetime of the stored values. Another security mechanism is added by additional information to determine the agreed usage. The field itself contains information on all other people, roles, service identifier, etc. that are allowed to access the entry. Since a list of all people or names that might access the entries is unfeasible the NIM can be used by a system using role based access control (RBAC) concepts. The role for users or a group of user can be added to the agreed usage data field of an entry.

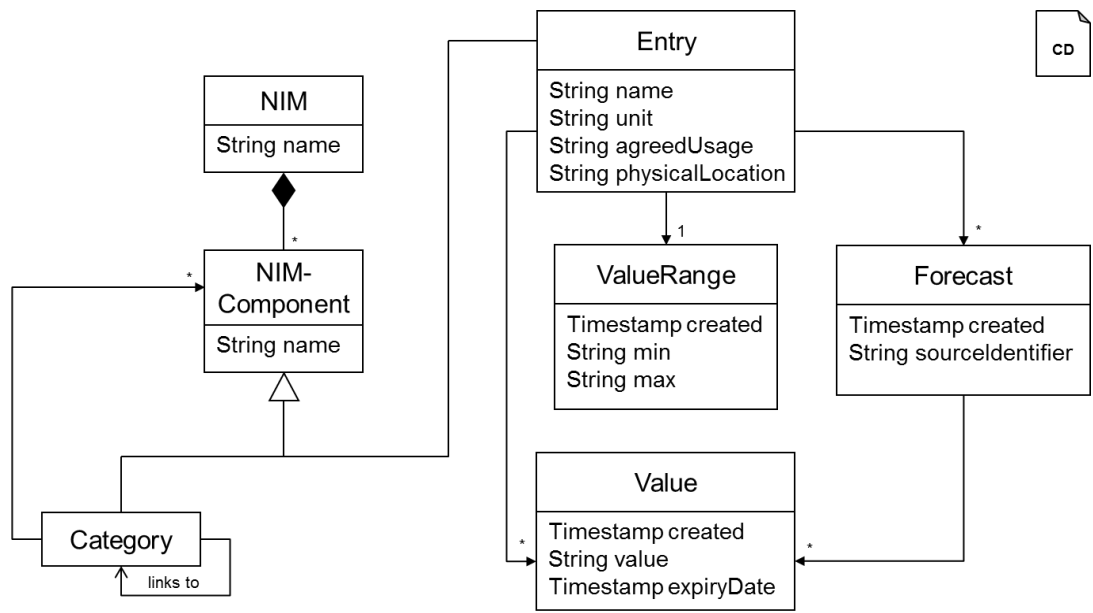

Figure 2. A UML class diagram detailing the generic neighbourhood information model. The two possible NIM-Components are shown as well as nesting possibility of components to enable hierarchy.

The last security aspect is the specification of the geographical, physical location where entries or values are allowed to be store. This information is necessary since most countries have different laws on privacy and users may want to decide in which countries or geographical locations entries and values of the NIM should be stored. The information can then be used by an implementation to determine the storage location. Moreover this information can serve as a decision base for users when deciding which information should be uploaded to a platform which uses the NIM as data format.

### 3.3 Implementation of the Neighbourhood Information Model

When technically implementing such a NIM on a platform additional aspects have to be taken into account. Especially the frequency of new values has to be considered. Automated and regularly measured values which are entered into such a platform can lead to a high data load which has to be handled in a different way than structural data of the neighbourhood which only changes sporadically over time. This can be achieved by e.g. only sending messages when there is a change in the value so that the last sent message value remains the current or real-time value. There are several reasons for this decision, but the main reasons should be performance and efficient use of storage capacity.

Also other important non-functional requirements have to be taken into account. Since two different neighbourhoods mostly do not have exactly the same data models, and future neighbourhoods may need data which cannot be foreseen at the moment or a single neighbourhood may evolve over time and thus, the corresponding data model has to be adapted, an adaptable and extensible implementation of the NIM is necessary.

To provide an adaptable and extensible implementation a mechanism for storing unforeseen entries that may arise during runtime of the overall systems is necessary. Therefore the prototype makes use of the generic meta model as its internal domain model which allows for the handling and persistence of arbitrary data entries fitting the meta model. Nevertheless the entries from a different data model have to be transformed into the meta model to be stored in the database. However this approach has the drawback that it works on the basis of a very general domain model which is uncomfortable to use. Moreover, data entries from other data models have to be transformed into the meta model format which would lead to enormous effort if implemented manually for each existing data model. But even if such a transformation would available via a manual implementation, the extension at runtime would not be possible for the new NIM entries. To overcome these problems we make use of a plugin based architecture that allows us to add new plugins at runtime. The plugins themselves are generated by a code generator that processes our input models at runtime. The generated code includes code responsible for the transformation between data in the concrete and generic format. In this way both problems have been solved.

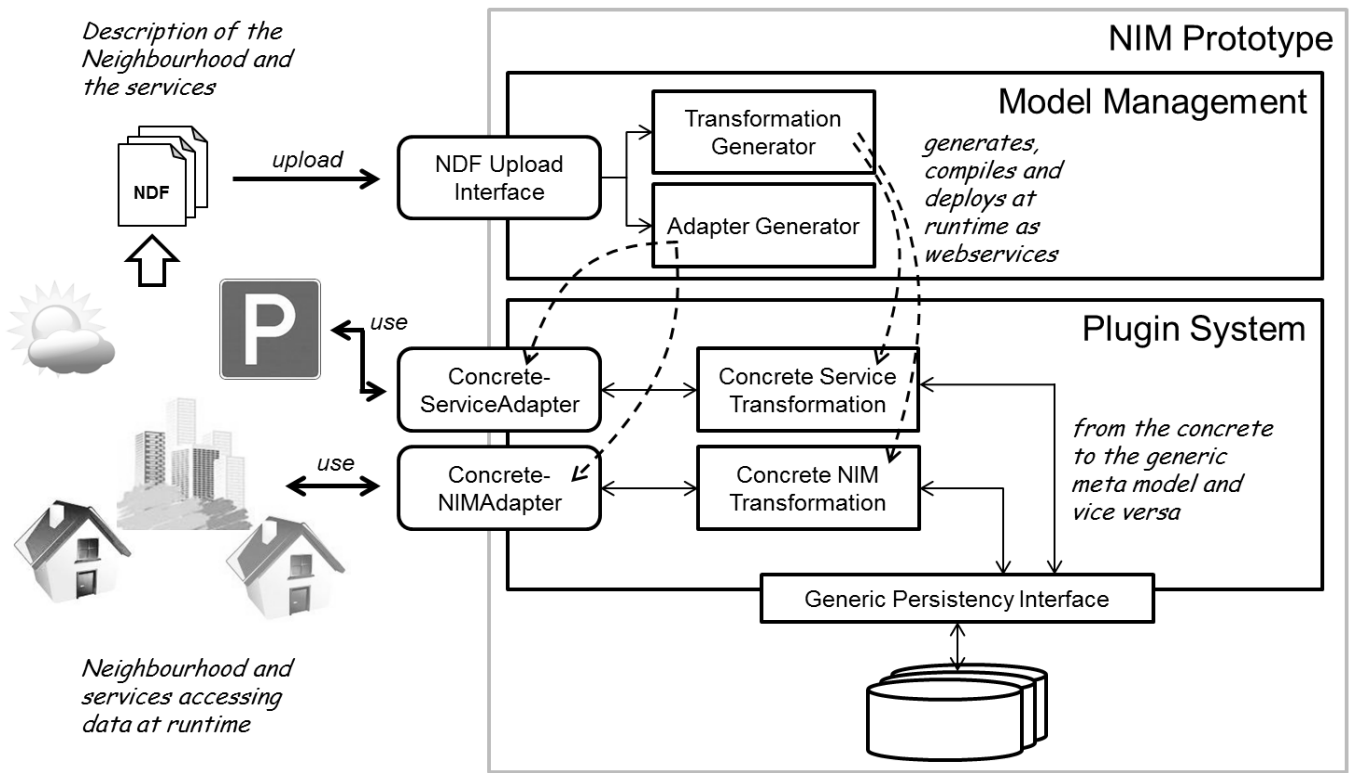

Figure 3. Overview of the overall implementation showing the Model Management component responsible for generating the adapters and the transformations, as well as the Plugin Systems being the runtime environment of the adapters.

The models are written in a DSL invented for describing concrete NIM data models in a NIM data format (NDF). It should be noted that the NDF specifies a data model and not concrete data. The methodology to configure the running prototype is shown in Figure 3.

The implementation consists of two components: a Model Management component and the Plugin System. The Model Management component is responsible to generate new adapters and transformation code if a new model is uploaded. The Plugin System itself represents the overall system that is used by services, and neighbourhood elements for accessing the data. To connect a new neighbourhood or a single service to the platform three steps are necessary.

In the first step an NDF model, specifying the concrete NIM format, has to be written in a textual syntax and uploaded to the prototype. The Model Management component analyses the NDF, checks context conditions and if it is valid, the component internally passes the model to the transformation generator and the adapter generator which generate a concrete adapter and transformation services. Both parts are then deployed to the Plugin System automatically. It should be mentioned that both the code generation and the deployment of the adapter and the transformation are performed at runtime of the system. The adapter serves as a webservice endpoint and offers an interface related to the concrete data model. It enables storing and retrieving the concrete data types. The transformation contains the logic on how data from a data model can be transformed to the generic data model and vice versa, which will be discussed after giving an example of the NDF.

The second step is then the usage of the new adapter by neighbourhood services operating on the defined data Model. The NDF models allow specifying the structure of neighbourhood information data. For example we could specify that rooms have a room name like this:

```
Room {
  String roomName;
}
```

This example clarifies that the NDF specifies a data format and not the concrete data, since we specified that each room has a room name, but not how many rooms exist and how the concrete names of the rooms are.

Since a single NDF file for the whole data model may become unfeasible and different neighbourhood services may need a distinct set of data entries of the concrete NIM, the definition of several data models in several NDFs is possible. Thus also different

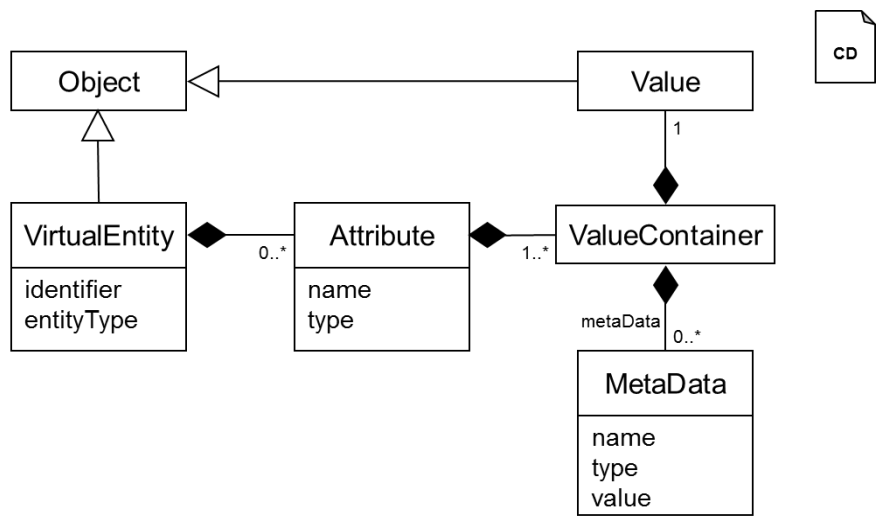

Figure 4. The edited domain model of the Internet of Things (IoT) approach (IoT.est, 2013)

stakeholders as well as specific service developer may develop their own description. Consider two different buildings with different data models, should be connected to the Neighbourhood Information System. In this case a different definition of a room, for each building can be provided. In combination with the previous example such a second extension could look like:

*AnotherRoom {*
 *String roomID;*
 *Number surface;*
*}*

Regarding the transformation between the concrete data model and the generic one we have to consider several cases. The first case, transforming a concrete data model into the generic data model is relatively straight forward. Taking up the example of the *Room*, this would become a category with name "room" within the generic data model. Additionally the category room would have an entry with name *roomName*. Since the NDF only describes data types and no instances the values of the entry are allocated at runtime and would contain the actual room name. Furthermore at runtime there would possibly be several instances of a room. The DSL also allows for hierarchically defining data types within other data types. This can be reflected via the category composite in the generic data model.

The second case is the transformation from the generic data model back to the concrete data model. If the data stored in the generic model originates from the same concrete model it should now be transformed back to, the transformation is as straight forward as the previously explained transformation to the generic data model.

The third case is the transformation of data originating from a different concrete model than it should be now transformed to. It is most likely that this is the most common case since every service and neighbourhood element wants to access data from other systems and has to know their data model therefore. Assuming that there may be an additional service which needs information about all buildings independent of the format, there are two possibilities:

- Using all the required existing adapters separately
- Specifying an additional NDF with a mapping included

Of course a service implementation can use every existing adapter separately and aggregate the data manually from the different formats. Nevertheless this is not envisioned and unfeasible.

The second alternative is more comfortable. The service developer can define the data in his own format like:

*StandardRoom {*
 *String identifier;*
*}*

In addition he needs to specify the mapping between his own room type and the already existing types:

*StandardRoom.identifier := Room.roomName |*
 *AnotherRoom.roomID;*

This type mapping information can be added in addition to the format definition in the same NDF model file. Note that the data type name serves as a namespace and the included field can be directly accessed. The NDF also contains a package declaration concept, not illustrated here, that enlarges the possible namespace and thus name clashes are to be considered unlikely. The list of values that are used for the identifier is separated by "|" and the list elements reference specific field from other NDFs. Additionally context condition checks ensure that the referenced types exist. On a technical level this is achieved by the language integration concepts of-fered by MontiCore (Völkel, 2011) (Look, et al., 2012). The generated adapter of our implementation takes care of the aggregation and offers the following methods to access the new defined Data:

*List<StandardRoom> getAllStandardRooms();*

This method allows resolving the data of all rooms, automatically converted in the standard room format.

## 4 RELATED WORK

In this section we provide work related to the NIM and to the integration of heterogeneous data sources.

In general there are two types of related work to be considered: Those following a standardization approach and those following a generic approach. Both differ in the way of integrating different heterogeneous data sources. Even within the two approaches there are different techniques applied for matching one data model to another. This can be done by ontologies, link models or adaptation.

Additionally approaches focusing on static data provided by information models and approaches focusing on measured, calculated building management systems (BMS) data have to be distinguished. Approaches like the OPC Unified Architecture (OPC UA), as an IEC standard (IEC 62541) (International Electrotechnical Commission, 2008), already provide an abstraction from the various management systems and will not be discussed in more detail. Nevertheless for static data there exist approaches such as the Industry Foundation Classes (IFC), registered under ISO-16739 ( buildingSMART International Limited, 2013), the Building Information Model (BIM) (Eastman, et al., 2011), CityGML (Open Geospatial Consortium, 2007) and gbXML (gbXML, 2013), just to name a few. While most of these approaches stop at the building envelope they also do not provide extension points for including additional data. Within CityGML it is possible to define extensions but these extensions also rely on a generic model. A detailed discussion of the advantages and disadvantages can be found within the Adapt4EE project (Tzovaras, et al., 2012).

Approaches like the HESMOS project (Hesmos Project, 2012) are trying to setup an integration platform that can integrate different data models via link models (Liebich, et al., 2011). These link models basically serve as adapters between the different data types.

Within the eeSemantics Community (eeSemantics, 2012) there are several research projects aiming at solving this issue. Approaches, such as the SEMANCO project (Semanco project, 2012) aim at defining a standardized data model as ontology, which is able to capture most of the data occurring in a building or in a neighbourhood.

Ontologies are commonly used when deducting new information from existing data models or when defining common concepts or mappings between data models. The main difference to our approach is that ontologies are mainly applied between two concrete data models and are thus bound to the concepts provided by the models.

Within the SEMANCO project a comprehensive list of possible data has been identified. The community also aims at finding ontologies for other areas apart from building or neighbourhoods. Thus there exists the FIPA device ontology (Foundation for Intelligent Physical Agents, 2002) and several smaller ontologies created within the AMIGO project (Georgantas, 2006). There are also concrete approaches considering sensor ontologies (Barnaghi, et al., 2012), linked data (Le Phuoc, et al., 2011) and sensors as a service (Compton, M., et al. , 2012) existent. Nevertheless, following a service oriented approach within the neighbourhood, this fixed data model is too rigid to be able to capture newly arising data created by services running within the neighbourhood. The underlying data model has to be changed if new information does not fit the current model. Thus our approach relies on a meta model backend and a mapping that is quite similar to ontology mapping approaches (Euzenat & Shvaiko, 2007). Other approaches like the Internet of Things approach (IoT-A) (IoT.est, 2013) (OpenIoT, 2013) (Probe-IT, 2013) already rely on a generic data model, shown in Figure 4.

Our Implementation of the NIM is close to the IoT-A model where we chose to have no explicit virtual elements and no explicit metadata. Nevertheless our model is not restricted by this.

## 5 CONCLUSION

Within this paper we have presented our approach to integrating heterogeneous building and periphery data models into a neighbourhood information model. We have shown how the mapping from a concrete manifestation of a data model to the underlying data model can be achieved and how the domain expert is enabled to express domain knowledge in a DSL which in turn is used to fully generate the required code at runtime of the overall system.

Nevertheless for the future we envision more work on the tooling infrastructure. At the moment the service developer has to know the existing data models and has to extend the mapping if new data models are integrated into the system. We envision this to be resolved by a tooling infrastructure providing access to a library of existing models.

Moreover the mapping defined between different models does not allow to distinct between specific instances. It is not possible to receive the data from specific instance in the standardized format, yet. For now, the information of all available instances matching the mapping instruction is resolved.

# 6 ACKNOWLEDGEMENTS

The work presented in this paper has been carried out in the COOPERaTE project, co-funded by the European Commission within the 7[th] Framework Programme (FP7/2007-2013) under grant agreement no 600063.